\documentclass[%
reprint,
 amsmath,amssymb,
 aip,
]{revtex4-2}
\usepackage{graphicx}
\usepackage{dcolumn}
\usepackage{bm}
\usepackage{hyperref}
\hypersetup{
  colorlinks=true,
  linkcolor=blue,
  citecolor=blue,
  urlcolor=black,
}
\usepackage{xcolor} 
\usepackage{ulem}

\usepackage{tabularx}
\usepackage{indentfirst}
\usepackage{mathtools}

\newcommand{\ket}[1]{|{#1}\rangle}
\newcommand{\bra}[1]{\langle{#1}|}
\newcommand{\braket}[2]{\langle{#1}|{#2}\rangle}
\newcommand{\ketbra}[2]{|{#1}\rangle\langle{#2}|}

\newcommand{\CNOT}{\text{CNOT}}

\begin{document}

\title[]{A Novel Superconducting Two-Qubit System Based on Transmon and PPQ,\\ and the Design of a CNOT Gate via Cross-Resonance Pulses}

\author{Jeongsoo Kang}
  \email{jskang1202@hanyang.ac.kr}
\author{Younghun Kwon}%
 \email{yyhkwon@hanyang.ac.kr}
  \affiliation{%
  Department of Applied Physics, Hanyang University, Ansan, 15588, Kyunggi-Do, South Korea.
}%




\date{\today}

\begin{abstract}
The transmon, known for its fast operation time and coherence time on the order of tens of microseconds, is the most widely used qubit in superconducting quantum processors. However, for the practical implementation of fault-tolerant quantum computing, the coherence time and gate fidelity of superconducting quantum processors must be improved. A novel superconducting qubit that protects the Cooper pair parity on a superconducting island has been proposed, offering better coherence performance than the transmon. However, the most efficient method for realizing a superconducting heterogeneous system that leverages this qubit remains unknown.
In this work, we propose that a cross-resonance-based CNOT gate can be designed in such a system. We provide the theoretical hardware specifications and pulse parameters required to simulate a two-qubit gate in the heterogeneous system. Our numerical simulation predicts a CNOT gate fidelity of 0.9989. This heterogeneous system offers a promising new platform for quantum computing.

\end{abstract}

%
%
%
%
%

\maketitle

Superconducting qubits\cite{Bouchiat98,Nakamura99,Vion02,Koch07,Paik11,Larsen20,Smith20,Hyyppa22,Chiorescu03,Harris10,Larsen15,Devoret85,Orlando99,Majer07} are among the most promising platforms for quantum computing. Based on the Cooper-pair box (CPB)\cite{Nakamura99,Bouchiat98}—which consists of a Josephson junction\cite{Josephson62,Josephson64} shunted by a capacitor—various designs have evolved, including charge qubits and transmons. The transmon\cite{Koch07,Majer07}, currently the most widely used superconducting qubit, along with flux qubits\cite{Orlando99,Chiorescu03,Harris10}, offers sub-microsecond gate times and greater resilience to background charge noise compared to charge qubits\cite{Nakamura02,Astafiev04,Schreier08,Schlor19}. This improved noise tolerance stems from the transmon’s large shunt capacitor, which reduces sensitivity to charge fluctuations\cite{Schreier08}.

Despite these advances, transmon-based quantum computers still struggle to execute large-scale quantum algorithms.
One of the factors that greatly affects the performance of these quantum processors is gate fidelity. Gate fidelity is reduced due to incoherent errors such as relaxation and dephasing, as well as coherent errors based on the structural limitations of the quantum processor. To overcome these challenges, researchers have explored using alternative superconducting materials\cite{Place21,Wang22} and optimizing processor architecture\cite{Kang24} to build logical qubits\cite{Kim23,Acharya24}.


Recently, a parity-protected superconducting qubit (PPQ) has been proposed\cite{Larsen20,Smith20}, which allows two Cooper pairs to tunnel simultaneously onto a superconducting island. The key principle is parity protection via a tunable potential well. External voltage\cite{Larsen20} or magnetic flux\cite{Smith20} can switch the well between single- and double-well configurations. In the double-well regime, eigenstates are distinguished by its parity of Cooper pairs and not allowed to transit to states in the other parity. This architecture effectively preserves parity, resulting in significantly longer relaxation times than those of transmons\cite{Larsen20,Smith20}. Such long relaxation times are particularly advantageous for executing deep quantum circuits, motivating the design of PPQ-based quantum processors\cite{Chirolli22}.

Despite PPQ's promising coherence properties, modeling a heterogeneous superconducting system that incorporates PPQ and designing operations for the heterogeneous system remains an open challenge.
A recent study demonstrated a controlled-phase gate between a transmon and a PPQ coupled via a capacitor\cite{Maiani22}, using a flux pulse to modulate the tunable transmon’s frequency.
This research highlights the potential of heterogeneous systems combining transmons and PPQs. However, capacitive coupling has limitations in scalability and spatial layout. To address this in homogeneous transmon systems, solutions including tunable transmon\cite{Caldwell18,DiCarlo09,Garcia20} such as floating tunable couplers\cite{Sete21} and waveguide extenders\cite{Marxer23} have been proposed. These technologies could enable long-distance flux-based two-qubit gates, but further investigation is needed to determine whether they can support large-scale superconducting processors.


Microwave voltage pulses, in contrast, are widely used in superconducting quantum platforms because they do not directly affect qubit frequencies and can be applied via direct connections to the qubit islands. A CR pulse\cite{Sheldon16,Kirchhoff18,Rigetti10} enables high-fidelity CNOT gates based on voltage pulses. To implement this, the two qubits have similar structures and are slightly detuned\cite{Rigetti10,Magesan20}. However, it has been unclear whether CR-based CNOT gates could function in heterogeneous systems like those involving PPQs, due to large frequency mismatches between transmons and PPQs.

In this work, we theoretically propose a method to perform a two-qubit gate based on a CR pulse in a novel heterogeneous system comprising a tunable transmon and a PPQ. The two qubits are indirectly coupled through a resonator, making careful theoretical design of the hardware parameters essential for the CR-based CNOT gate to function properly. We provide the optimal hardware and pulse parameters necessary to simulate this CNOT gate. Based on these parameters, our numerical results yield a CNOT gate fidelity of approximately 0.9989. This suggests that the transmon-PPQ system lays a new foundation for advancing toward heterogeneous superconducting quantum computing.


\begin{figure*}[t]
  \centering
  \includegraphics[width=1.0\linewidth]{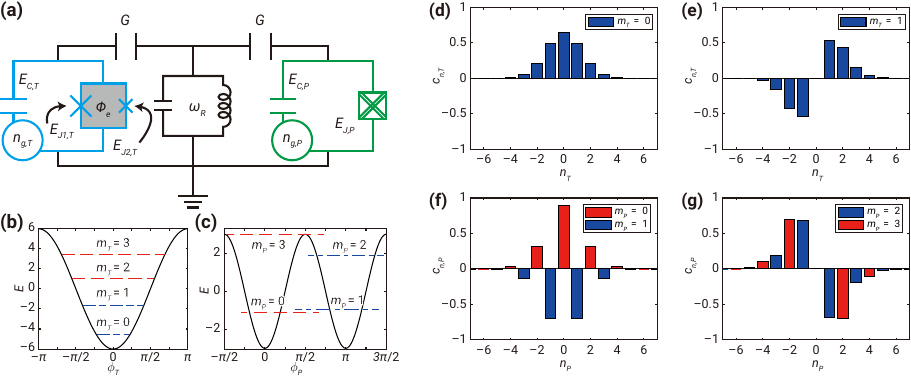}
  \caption{
    Design of the transmon-PPQ system and eigenstates of the transmon and the PPQ, comprising the transmon-PPQ system.
    (a)Lumped-element circuit diagram of the transmon-PPQ system. In the system, a tunable transmon qubit(blue) and a PPQ(green) are connected through a resonator.
    The Josephson energy depends on $E_{J1,T}$ and $E_{J2,T}$, which are the Josephson energy of two Josephson junctions, and the external magnetic flux $\Phi_e$ through the DC-SQUID structure (gray area).
    The connection between the resonator and each qubit is a capacitive connection with the coupling strength $G$.
    The transmon and PPQ can be driven by their offset numbers $n_{g,T}$ and $n_{g,P}$, respectively.
    Energy levels of (b) the idle transmon and (c) the PPQ. Blue(red) dashed lines indicate the energy levels corresponding to the computational basis(leakage).
    We depict the eigenstates in the charge representation to compare the Cooper-pair parity in the island of each qubit. Eigenfunctions of the transmon, when (d) $c_{n,T}=\braket{n_T}{m_T=0}$ and (e) $\braket{n_T}{m_T=1}$, respectively. The eigenfunctions of the PPQ, where (f) $c_{n,P}=\braket{n_P}{m_P=0,1}$ and (g) $\braket{n_P}{m_P=2,3}$.
  }
  \label{fig:Tr-PPQ_system}
\end{figure*}

We consider a transmon-PPQ system composed of a tunable transmon qubit and a PPQ (Figure \ref{fig:Tr-PPQ_system}a). The tunable transmon and the parity-protected superconducting system are connected through a coplanar waveguide resonator. The energy levels of the transmon and PPQ are shown in Figure \ref{fig:Tr-PPQ_system}b and \ref{fig:Tr-PPQ_system}c, respectively. Detailed explanation of Hamiltonian and hardware parameters of the transmon-PPQ system are in Supplementary Materials.

For the transmon qubit, a Cooper-pair penetrates the effective Josephson junction, an asymmetric DC-SQUID loop. This suggests that each eigenstate is distinguished by its plasmon mode, and not by the parity of the Cooper pair on the island.
Here, plasmon means oscillation of quantized charge across the inductive loop and the shunt capacitance\cite{Smith20}. For instance, plasmon in a transmon is oscillation of the number of Cooper pairs in a superconducting island. The ground(first excited) mode of the plasmon in the transmon is illustrated in Figure \ref{fig:Tr-PPQ_system}d(e). One can easily notice that the plasmon modes determine the envelope of the wavefunctions that represent the eigenstates.

The frequency tunability of the transmon qubit is not utilized for qubit operations. However, tunability can be used to control fluctuations in hardware specifications, thus compensating for the long-term drift that occurs in superconducting quantum processors. We design two-qubit gates using microwave pulses in a transmon-PPQ system. An important step in designing such gates is understanding the qubit frequency configuration. Thus, we employ a transmon which has a frequency tunability as a new building block.

In contrast to the transmon, only a pair of Cooper-pairs are allowed to access the island. Thus, the eigenstates of the PPQ can be distinguished by the parity of the extra Cooper-pair on the island, even if the plasmon modes are identical (Figure \ref{fig:Tr-PPQ_system}f and \ref{fig:Tr-PPQ_system}g).
The transition between two states with the same parity but different plasmon modes in the PPQ is achieved by using microwave pulses, similar to those used to control conventional transmon. In contrast, for transitions between two states within the same plasmon mode but with different parities, controlling the external flux on PPQ is a possible method. This method enables the PPQ to reveal the single Cooper-pair tunneling operator, but simultaneously make it vulnerable to flux noise. In this study, we select two plasmon states with the same parity as the computational basis.

To design a two-qubit gate based on the CR effect, we model the system such that the two qubits are connected via a coplanar waveguide resonator.

While the coupling strengths may differ in a real chip due to errors in the fabrication process, CNOT gates based on CR can be implemented if the coupling energies are in the order of tens of MHz.

One method of designing CNOT gates in a transmon-based quantum computer is to utilize a CR pulse. The CR effect is conditional resonance between two transmon qubits, in which an axis of induced Rabi oscillation in the target qubit flips according to the state of the control qubit. It is only known that the CNOT gate based on the CR pulse can be implemented in a homogeneous superconducting system, such as a transmon-based processor, where coplanar waveguides mediate qubit-qubit interactions\cite{Rigetti10,Magesan20}.

In this study, we theoretically propose that the CR effect can be induced in heterogeneous superconducting systems. As an example, we design a CNOT gate based on CR pulses implemented in a transmon-PPQ system.
We explain that the transmon-PPQ system reduced to the two-qubit model is equivalent to a homogeneous two-qubit system at the Hamiltonian level. Moreover, we present numerical results indicating that CR pulses can operate successfully even when the structure contains higher levels.

It is well-known that a transmon can be modeled as a quantum Duffing oscillator\cite{Koch07,Bishop10}. The uncertainty principle between the number and the phase of Cooper pairs, $[\hat{\phi}_T,\hat{n}_T]=i$, allows us to express the Hamiltonian of the transmon using the second quantization of $\hat{n}_T \propto (\hat{b}_T+\hat{b}_T^\dagger)$, where $\hat{b}_T$ is lowering operator of the transmon. This denotes that the Hamiltonian of the transmon reduced to TLS is written as,
\begin{equation}
  \hat{H}_{T,\text{TLS}} = \omega_T \ketbra{m_T=1}{m_T=1} + \Omega_T^\text{TLS}(t) \hat{\sigma}_{x,T},
\end{equation}
where the transverse driving source $\Omega_T^\text{TLS}(t) = -8E_{C,T}n_{g,T}(t)$ and the Pauli X operator of the transmon $\hat{\sigma}_{x,T}$ is defined as $\ketbra{m_T=0}{m_T=1} + \ketbra{m_T=1}{m_T=0}$.

We show that a PPQ reduced to TLS has the same analogy of a transmon qubit.
It is known that the charge operator can distinguish the parity of the Cooper pair on the island of the PPQ\cite{Aasen16} and, for $\cos 2\hat{\phi}$ devices, opposite-parity states are decoupled\cite{Smith20}.
It implies that the dynamics of the PPQ, where the parity is fixed, can be rewritten with the conjugate observables of a pair of Cooper pairs, $\hat{\mathcal{N}}_P=\hat{n}_P/2$ and $\hat{\delta}_P=2\hat{\phi}_P$.
We only focus on the odd-parity states $\ket{m_P=1}$ and $\ket{m_P=2}$ in this work. Then, likewise the transmon, the dynamics of the one-sided parity can be rendered as a weakly anharmonic oscillator,
\begin{equation}
    \hat{H}_P \simeq \tilde{\omega}_P \hat{b}_P^\dagger \hat{b}_P - 2E_{C,P}\hat{b}_P^\dagger\hat{b}_P\left(\hat{b}_P^\dagger\hat{b}_P - 1\right),
\end{equation}
where $\tilde{\omega}_P=\sqrt{32E_{C,P}E_{J,P}}-4E_{C,P}$ and $\hat{b}_P$ is the lowering operator of the PPQ.
Thus, we can apply the same approach to comprehend the PPQ as we use to understand the transmon qubit.
In the context of TLS, $\hat{\mathcal{N}}_P \propto (\hat{b}+\hat{b}^\dagger)$ can be reduced to $\hat{\sigma}_{x,P} = \ketbra{m_P=1}{m_P=2} + \ketbra{m_P=2}{m_P=1}$. 
Then, the PPQ has the same analogy of a two-level atom driven by external electric field\cite{Blais04} as following:
\begin{equation}
  \hat{H}_{P,\text{TLS}} = \tilde{\omega}_P \ketbra{m_P=2}{m_P=2} + \Omega_P^\text{TLS}(t) \hat{\sigma}_{x,P},
\end{equation}
where $\Omega_P^\text{TLS}(t) = -8E_{C,P}n_{g,P}(t)$. We note that the qubit frequency of the PPQ, $\omega_P$, can be different from the approximated frequency $\tilde{\omega}_P$ in our system. It is because $E_{J,P}$ is not large enough to enter the region $E_{C,P} \ll E_{J,P}$. Nevertheless, the dynamics of the PPQ still holds the same analogy of a anharmonic oscillator.

Then, we couple these two Hamiltonians by adding a resonator and qubit-resonator interaction Hamiltonians. Transverse two-qubit interaction, such as $XX$ or $YY$, is needed to operate a CR gate. This type of interaction can be simply achieved by a capacitive coupling between two superconducting qubits, where the superconducting islands are close together to interact directly or connected to a quantum bus. To reflect the practical need for spatial freedom in physical architectures, we model our system using the latter method.
As we perform Schrieffer-Wolff transformation\cite{Schrieffer66} to the transmon-PPQ system, a reduced TLS Hamiltonian $\hat{H}_\text{TLS}$ is approximated as,
\begin{equation}
  \hat{H}_\text{TLS} =
  \hat{H}_{T,\text{TLS}} + \hat{H}_{P,\text{TLS}} +
  J \hat{\sigma}_{x,T} \hat{\sigma}_{x,P},
\end{equation}
where,
\begin{equation}
  J = \frac{G^2(\bar{\omega}_T + \bar{\omega}_P - 2 \omega_R)}{2(\bar{\omega}_T-\omega_R)(\bar{\omega}_P-\omega_R)}.
\end{equation}
Bare frequencies of the transmon(PPQ) $\bar{\omega}_{T(P)}$ is set as 2.8827(2.8467) GHz. The derivation of $J$ is based on the work of E. Magesan\cite{Magesan20}.

Finally, We set the qubit frequencies of the transmon and the PPQ close enough. We present one of the possible hardware parameters. In this case, the qubit frequency of the transmon, $\omega_T$, is 2.8830 GHz and those of the PPQ, $\omega_P$, is 2.8470 GHz. Then, the resonator, which has the resonance frequency $\omega_R$ of 2.4 GHz couples these qubits. The capacitive coupling strength between the qubits and the coupler, $G$, is set as 10 MHz.
The effective Hamiltonian of CR pulse\cite{Magesan20} in the TLS is given as,
\begin{equation}
  \hat{H}_\text{eff,CR} = \left(\Delta - L\right) \frac{\hat{\sigma}_{z,T}\hat{I}_P}{2} - \left(\frac{J\Omega}{L}\right)\frac{\hat{\sigma}_{z,T}\hat{\sigma}_{x,P}}{2},
  \label{eq:CR_Heff}
\end{equation}
where $\Delta = {\omega}_T - {\omega}_P$ and $L = \sqrt{\Delta^2 + \Omega^2}$. Here, the driving pulse is considered as a monochrome microwave pulse, $\Omega^\text{TLS}(t) = \Omega \cos(2\pi f t)$.
Considering the order of parameters in Eq.\ref{eq:CR_Heff}, this configuration allows the transmon-PPQ system to interact using a CR pulse without modal interactions, i.e., spontaneous energy exchange due to the resonance.

In this section, we present how to design quantum gates in the transmon-PPQ system. For single-qubit gates, it is straightforward to design them in each qubit by applying microwave pulses. We explain the design of single-qubit gates in Supplementary Materials.
We focus on designing two-qubit gates, specifically CNOT gates, based on the CR effect in the transmon-PPQ system.
\begin{figure}[t]
  \centering
  \includegraphics[width=0.9\linewidth]{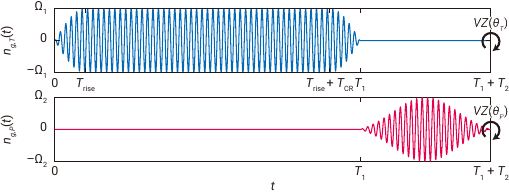}
  \caption{
      The proposed pulse design of $\CNOT_\text{TP}$ gate.
      The sequence starts with applying the CR pulse to the transmon.
      After applying the CR pulse $n_{g,T}(t)$ to control qubit, an auxiliary pulse $n_{g,P}(t)$ is performed on the target qubit. Here, the transmon is the control qubit and the PPQ is the target qubit. The CR pulse(auxiliary pulse) is characterized by pulse parameter vector $\mathbf{x}_\text{CR}$($\mathbf{x}_\text{Aux}$).
      The final step adjusts the phase information of each qubit by applying the VZ
      gates.
  }
  \label{fig:pulse_protocol}
\end{figure}
\begin{figure*}[t]
  \centering
  \includegraphics[width=0.9\linewidth]{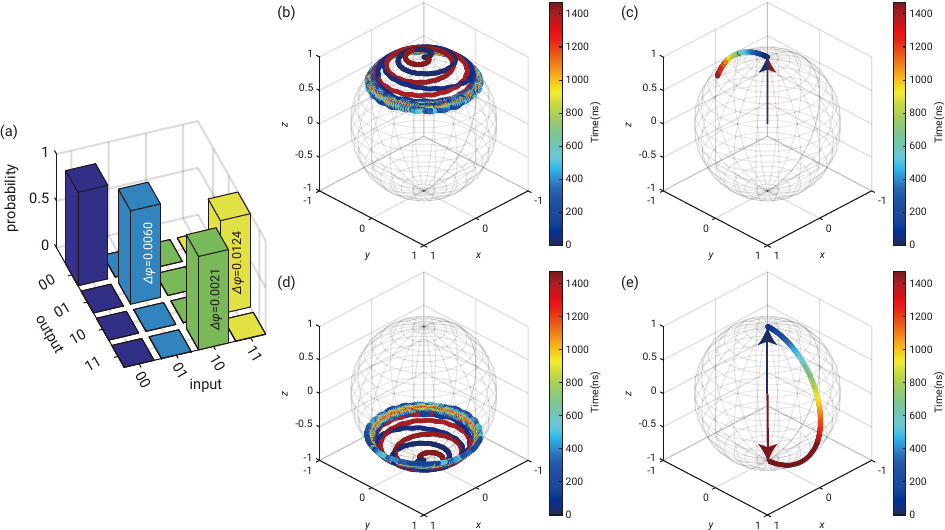}
  \caption{
    Numerical probabilities and the trajectory of the Bloch vectors for the basis states after applying the $\CNOT_\text{TP}$ gate. We highlight the initial(blue) and the final(red) states of the target qubit as the colored arrows.
    (a) The simulated state tomography of $\CNOT_\text{TP}$ for the four basis states. $\Delta\varphi$ denotes the relative phase for the $\ket{00}$ state.
    The trajectory of the Bloch vectors for (b)the transmon and (c)the PPQ during the $\CNOT_\text{TP}$. The initial state for the demonstration is $\ket{00}$.
    The trajectory of the Bloch vectors for (e)the transmon and (f)the PPQ during the $\CNOT_\text{TP}$. The initial state for the demonstration is $\ket{10}$.
  }
  \label{fig:2Q_Bloch}
\end{figure*}
\begin{table*}[t]
  \centering
  \caption{
      The pulse parameters of $\CNOT_\text{TP}$ gate and the average fidelity $F$ in transmon-PPQ system. Here, $\theta_1 = 0.6007$ and $\theta_2=-0.0333$. All the parameters are optimized to maximize the average fidelity by using the Nelder-Mead algorithm\cite{Nelder65}. See Supplementary Materials for the details of optimization.
  }
  \label{table:CNOT_pulse_parameter}
  \begin{tabular*}{\textwidth}{@{\extracolsep{\fill}}cccccc}
  \hline
  Gate                 & $f_1$  & $f_2$  & $T_1$  & $T_2$  & $\Omega_S$ \\ \hline
  $\CNOT_{\rm TP \it}$ & 2.8470 & 2.8472 & 1460.0 & 9.9966 & 0.03000    \\ \hline \hline
  $\Omega_G$ & $\varrho$ & $\gamma_1$ & $\gamma_2$ & $(\theta_T,\theta_P)$ & $F$    \\ \hline
  0.02078    & 0.09986   & -1.068E-06 & 2.4186     & $(\theta_1, \theta_2)$   & 0.9989 \\ \hline 
  \end{tabular*}
\end{table*}

The CR pulse applied to the control qubit is a microwave pulse whose frequency is approximately equal to that of the target qubit. The CR pulse induces conditional resonance to rotate the target qubit state according to the control qubit state. Figure \ref{fig:pulse_protocol} illustrates the proposed pulse sequence for the CNOT gate based on the CR pulse in the transmon-PPQ system. The control(target) qubit is the transmon qubit(PPQ). The CNOT gate is represented by $\CNOT_\text{TP}$. The charge offset functions $n_{g,T}(t)$ and $n_{g,P}(t)$ in Figure \ref{fig:pulse_protocol} are defined as follows:
\begin{align}
  \label{eq:CR_protocol}
  n_{g,T}(t) &= \Omega_S(t) \cos(\omega_1 t - \gamma_1), \\
  n_{g,P}(t)  &= \Omega_G(t-T_1) \cos(\omega_2 t - \gamma_2).
\end{align}
The microwave pulse $n_{g,T}(t)$ designed for the transmon carries information regarding the CR pulse. The frequency of the CR pulse, $f_1=\omega_1/2\pi$, is set to be near the qubit frequency of the PPQ. The initial phase of the CR pulse, $\gamma_1$, determines the axis of the rotation of the target qubit state due to the CR pulse. We set the initial phase to zero to induce $X$ rotation on the target qubit.
The envelope of the CR pulse, $\Omega_S(t)$, is a sinusoidal flat-top function,
\begin{widetext}
  \begin{equation}
    \Omega_S(t) = \left\{\begin{array}{ll}
    \Omega_S \sin(\pi t/ 2T_\text{rise}) & t \in [0, T_\text{rise})\\
    \Omega_S & t \in [T_\text{rise}, T_\text{rise} + T_\text{CR})\\
    \Omega_S \sin\{\pi (t-T_\text{CR})/ 2T_\text{rise}\} & t \in [T_\text{rise} + T_\text{CR}, T_S = 2T_\text{rise} + T_\text{CR})\\
    0 & \text{Otherwise}
    \end{array}\right.
    .
  \end{equation}
\end{widetext}
The other pulse $n_{g,P}(t)$, namely the auxiliary pulse, is designed for the PPQ to complete the CNOT operation. The auxiliary pulse is a single-qubit gate $R_{X,P}$. Note that $\gamma_2$ can be set as non-zero to compensate time offset $T_1$. In Figure \ref{fig:pulse_protocol}, the rotating arrows at the time $t=T_1+T_2$ indicate the compensating Z rotations, VZ gates, with the rotating angle $\theta_i$ for each qubit.

We design the pulse sequence for the CNOT gate stepwise. The CNOT gate is decomposed into unitary operators corresponding to the pulse protocols $\hat{\mathcal{U}}$ and VZ gate $\hat{\mathcal{Z}}$,
\begin{equation}
  \CNOT_\text{TP} =
  \hat{\mathcal{Z}}\hat{\mathcal{U}}_\text{Aux}\hat{\mathcal{U}}_\text{CR}.
\end{equation}
$\hat{\mathcal{U}}_\text{CR(Aux)}$ is a time-evolution operator that illustrates how the transmon-PPQ system is driven by a CR(auxiliary) pulse. The time-evolution operator for the CR pulse depends on the pulse parameter vector $\mathbf{x}_\text{CR}=(f_1,T_S,\varrho,\Omega_S,\gamma_1)^T$, where $\varrho$ is the ratio of the rising time for the CR pulse. We investigate an appropriate pulse parameter vector that can induce an effective CR. We also note that finding proper hardware parameters is important for the PPQ to engage with the CR pulse. In summary, we set the parameters of the transmon-PPQ system before searching for the pulse parameter vector of the CR pulse.
The pulse parameter vector for the auxiliary pulse is given as $\mathbf{x}_\text{Aux} = (f_2\simeq\omega_2/2\pi,T_G,\sigma,\Omega_G,\gamma_2)^T$, where the thickness of the Gaussian envelope $\sigma$ is set as $T_G/4$. As the auxiliary pulse induces a single-qubit rotation of the PPQ, the pulse frequency is the same as that of the PPQ. We note that the initial phase, $\gamma_2$, plays a key role in increasing the gate fidelity of the CNOT. The initial phase determines the axis of rotation of the Bloch vector due to the auxiliary pulse. We frame the CNOT gate with an auxiliary pulse, rectifying the conditional rotation caused by CR. The VZ gate, $\hat{\mathcal{Z}} = \hat{I}^{(R)}\otimes \hat{R}_Z^{(T)}(\theta_T)\otimes \hat{R}_Z^{(P)}(\theta_P)$, is a frame rotation with the parameters $\theta_T$ and $\theta_P$, the rotating angles of each qubit.
Table \ref{table:CNOT_pulse_parameter} lists the optimized pulse parameters and the gate fidelity of $\CNOT_\text{TP}$. We emphasize that the CNOT gate with an average fidelity of 0.9989 is simulated in a superconducting heterogeneous system containing a transmon and PPQ. It suggests that high-fidelity two-qubit gates can be theoretically achieved in a heterogeneous superconducting qubit system.

Figure \ref{fig:2Q_Bloch} illustrates the numerical result of $\CNOT_\text{TP}$. We present the numerical probabilities after the gate operation (Figure \ref{fig:2Q_Bloch}a) and evolution of the Bloch vectors during the gate operation (Figure \ref{fig:2Q_Bloch}b-\ref{fig:2Q_Bloch}e).
Figure \ref{fig:2Q_Bloch}a shows that, for the basis states, our simulation yields a high success probability of the CNOT gate whose relative phases are almost zero.
Figure \ref{fig:2Q_Bloch}b and \ref{fig:2Q_Bloch}d show the trajectories of the control qubit states,
which implies that the adiabatic process is designed properly. For the target qubit, the PPQ experiences $X$ or $-X$ rotations according to the control qubit state (Figure \ref{fig:2Q_Bloch}c and \ref{fig:2Q_Bloch}e).
Since the initial states are in the computational basis, the effect of the VZ gate cannot be visualized.

In this study, we proposed a superconducting heterogeneous building block consisting of a transmon qubit and a PPQ. In our novel building block, we simulated not only single-qubit gates but also a high-fidelity two-qubit gate using microwave drive lines, employed in conventional transmon quantum processors. We revealed the hardware parameters of the transmon-PPQ system and the pulse parameters to implement a CNOT gate based on CR. The gate fidelity of the CNOT gate reached 0.9989, which is almost sufficient for error suppression using a quantum error-correcting code. We suggest that such a heterogeneous superconducting quantum hardware can be employed to achieve the fault-tolerant quantum computing.

The coupling strength does not change over time in this building block. It is known that the tunability of the couplings not only implements high-fidelity two-qubit gates but also suppresses crosstalk errors due to the non-nearest neighboring qubits. We suggest that the adjustable couplings using tunable couplers also can be considered in the superconducting heterogeneous system.

This work is supported by the Basic Science Research Program through the National Research Foundation of Korea (NRF) funded by the Ministry of Education, Science and Technology (NRF2022R1F1A1064459) and Creation of the Quantum Information Science R$\&$D Ecosystem (Grant No. 2022M3H3A106307411) through the National Research Foundation of Korea (NRF) funded by the Korean government (Ministry of Science and ICT).

\section*{Author Declarations}
\subsection*{Conflict of interest}
The authors declare no conflict of interest.

\subsection*{Author Contributions}
Jeongsoo Kang: Conceptualization (equal); Formal analysis (equal); Investigation (equal); Methodology (equal); Software (equal); Visualization (equal); Writing - original draft (equal); Writing - review \& editing (equal).
Younghun Kwon: Conceptualization (equal); Formal analysis (equal); Investigation (equal); Methodology (equal); Writing - review \& editing (equal).

\section*{Data availability statement}
The data that support the findings of this study are openly available in transmon-PPQ-system-data at \url{https://github.com/JeongsooKang/transmon-PPQ-system-data}.

\clearpage
\onecolumngrid
\appendix
\begin{center}
  \textbf{\large Supplementary Materials: A Novel Superconducting Two-Qubit System Based on Transmon and PPQ, and the Design of a CNOT Gate via Cross-Resonance Pulses}
\end{center}

\section{Hamiltonian of the Transmon-PPQ System}
\begin{table*}[b]
  \centering
  \caption{
      Hardware parameters of the transmon-PPQ system used in this work.
  }
  \label{table:Tr-PPQ_system_parameter}
  \begin{tabular*}{\textwidth}{@{\extracolsep{\fill}}cccc}
  \hline
    & Resonator (R) & Transmon (T) & PPQ (P) \\ \hline
  Resonance frequency $\omega_R$ & $2\pi\times 2.4~\text{GHz}$ & - & - \\
  Qubit frequency $\omega_i$  & - & $2\pi\times 2.8830~\text{GHz}$ & $2\pi\times 2.8470~\text{GHz}$ \\
  Charging energy $E_{C,i}$ & - & $2\pi\times 0.2~\text{GHz}$ & $2\pi\times 0.2~\text{GHz}$ \\
  Josephson energy $E_{J,i}$ & - & $E_{J,T}(\Phi_e=0) = 2\pi\times 6.0~\text{GHz}$ & $2\pi\times 3~\text{GHz}$ \\
  Ratio of Josephson energies $\gamma$ & - & $1.01$ & - \\
  Coupling energy $G_i$ & - & $2\pi\times 0.01~\text{GHz}$ & $2\pi\times 0.01~\text{GHz}$ \\
  \hline
  \end{tabular*}
\end{table*}
The total Hamiltonian of the transmon-PPQ system can be described as
\begin{equation}
    \hat{H}(t) = \hat{H}_T(t) + \hat{H}_P(t) + \hat{H}_R + \hat{H}_I,
\end{equation}
where the Planck constant $\hbar=1$.
Here $\hat{H}_i (i=T,P,R)$ denotes the Hamiltonian of the transmon, the PPQ, and the resonator, respectively. $\hat{H}_I$ means the interaction Hamiltonian between the resonator and qubits.

The Hamiltonian of the tunable transmon $H_T$ becomes 
\begin{equation}
    \hat{H}_T(t) = 4E_{C,T}(\hat{n}_T - n_{g,T}(t))^2 - E_{J,T}(\varphi_e) \cos \hat{\phi}_T.
\end{equation}
$\hat{n}_T$ is the extra Cooper pair-number operator of the transmon. $\cos\hat{\phi}_T$ is a tunneling operator such that a Cooper-pair penetrates a Josephson junction.
$n_{g,T}$ is the offset number of the transmon, and denotes the microwave pulse acting on the transmon to construct the gate. $E_{C,T}$ is charging energy and $E_{J,T}(\varphi_e)$ is the Josephson energy expressed as a function of the reduced external magnetic flux  $\varphi_e = ({\pi}/{\Phi_0})\Phi_e$,
\begin{equation}
  E_{J,T}\left(\varphi_e\right)
  = E_{J\Sigma,T}\left(\cos^2(\varphi_e) +
  \frac{\gamma-1}{\gamma+1}\sin^2(\varphi_e)\right)^{1/2},
\end{equation}
where $\Phi_0$ denotes the magnetic flux quantum and $\Phi_e$ does the  external flux.

A PPQ is a superconducting qubit in which only a pair of Cooper pairs tunnel into an island of CPB. The Hamiltonian of the PPQ is defined as,
\begin{equation}
  \hat{H}_P(t) = 4E_{C,P}(\hat{n}_P - n_{g,P}(t))^2 - E_{J,P} \cos 2\hat{\phi}_P,
\end{equation}
where $\hat{n}_P$ is an extra Cooper-pair number operator of PPQ and $n_{g,P}(t)$ is the microwave pulse for the PPQ. The tunneling operator $\cos 2\hat{\phi}_P$ explains that only a pair of Cooper pairs can be exchanged in the superconducting island of the PPQ. $E_{C,P}$ and $E_{J,P}$ denote the charging and Josephson energies of the PPQ, respectively.

The Hamiltonian of a resonator $\hat{H}_R$ becomes $\omega_R \hat{a}^\dagger \hat{a}.$ $\hat{a}$($\hat{a}^\dagger$) is an annihilation(creation) operator of a resonator. $\omega_R$ denotes the resonance frequency of the resonator.

There are two interactions in the transmon-PPQ system; one between the resonator and transmon and the other between the resonator and PPQ. The following Hamiltonian $H_I$ describes these interactions:
\begin{equation}
    \hat{H}_I = G_T (\hat{a}+\hat{a}^\dagger)\hat{n}_T + G_P (\hat{a}+\hat{a}^\dagger)\hat{n}_P.
\end{equation}
$G_{T(P)} = G$ denotes the coupling energy between the resonator and transmon(PPQ)\cite{Kang24}. 

We define the finite dimensional Hilbert space $\mathcal{H}$ as follows:
\begin{equation}
  \mathcal{H} = \mathrm{Span} \left\{
    \ket{\Psi} = \ket{k}\ket{m_T}\ket{m_P} :
    k, m_{T,P} \in \{0,1,2,3\}
  \right\},
\end{equation}
where all devices in the transmon-PPQ system are described as four-level quantum systems.
$\ket{k}$ is the Fock state of resonator. $\ket{m_T}$ denotes the energy eigenstate of the idle transmon.
Here, the `idle' state refers to the condition in which there are no microwave or flux pulses, except for the flux bias used to stabilize the qubit frequency.
Likewise, $\ket{m_P}$ is the energy eigenstate of the idle PPQ.
The effective Hamiltonian of a CR resonance containing high-energy levels is known to describe sophisticated dynamics that are closer to CR gate implemented in a real quantum hardware than the effective Hamiltonian approximated by a two-level system (TLS)\cite{Magesan20}.
Thus, $k,m_T,m_P \in \{0,1,2,3\}$ are included for accurate simulation.
The computational basis of the transmon-PPQ system $\mathcal{C}$ can be defined as,
\begin{equation}
  \mathcal{C} = \mathrm{Span}\{
      \ket{ij}
      = \ket{k=0}\ket{i}\ket{j+1}:i,j\in\{0,1\}
  \}.
  \label{eq:basis_qubit}
\end{equation}
Here, we consider the case in which a resonator does not absorb energy. Therefore, in the computational basis, the resonator remains at the ground state.
The ground and first excited states of the transmon are chosen as the computational basis. We choose the two lowest states $\ket{m_P=1,2}$, that have the same parity but are in different plasmon modes, as the computational basis for the PPQ.
These two states can be treated as conventional basis of transmon qubit, so that we can manipulate these states by using microwave pulses.
The resonator is just a transmission line to transfer energy between the two qubits. In other words, if the resonator is measured in an excited state, it indicates that the system is malfunctioning.

The presence of high-energy level in the system must be considered when designing a high-fidelity two-qubit gate. High energy levels are necessary to accurately describe their dynamics, but can also be used to understand leakage error.
We include these leakages $\mathcal{L}=\mathcal{H}-\mathcal{C}$ to accurately simulate the dynamics of a real multi-level quantum system. The number of bounded levels in the transmon and PPQ is determined by the hardware parameters. Nevertheless, the four lowest levels are sufficient to simulate the CR effect.

\subsection{Matrix Representation}
We represent the Hamiltonian operators of the resonator, transmon, and PPQ.
The Hamiltonian of the resonator is expressed simply in the Fock state representation. The annihilation operator $\hat{a}$ written as a $4\times 4$ matrix is the main component of the resonator Hamiltonian,
\begin{equation}
  \hat{a} = \left[
    \begin{array}{cccc}
      0 & 1 & 0 & 0 \\
      0 & 0 & \sqrt{2} & 0 \\
      0 & 0 & 0 & \sqrt{3} \\
      0 & 0 & 0 & 0
    \end{array}
  \right].
\end{equation}

The transmon and PPQ have different tunneling operators; however, they can be described by the same analogy of CPB structure. Thus, we explain the Hamiltonian of the CPB in a matrix representation.
The operators that appear in the CPB Hamiltonian are the number operator of the Cooper pairs $\hat{n}=\hat{Q}/2e$ and the tunneling operator $\cos(\hat{\varphi}=2\pi\hat{\Phi}/\Phi_0)$. $\hat{Q}$ and $\hat{\Phi}$ denote the charge operator and the magnetic flux operator, respectively.
According to the uncertainty principle, $[\hat{\Phi},\hat{Q}]=i\hbar$, the CPB Hamiltonian can be written as a matrix represented by the eigenstate $\ket{n}$ of $\hat{n} = \sum_{n \in \mathbb{Z}}{n \ketbra{n}{n}}$.
Note that the tunneling operator $\cos\hat{\varphi} = (e^{i\hat{\varphi}} + e^{-i\hat{\varphi}})/2$ is expressed as $(\ketbra{n+1}{n}+\ketbra{n}{n+1})/2$ due to the penetration of a single Cooper pair. For the CPB, since $E_C \gg E_J$, the eigenstate of the CPB Hamiltonian is almost identical to $\ket{n}$. Thus, the charge basis is the best option to represent operators in the form of matrices. In contrast, $\ket{n}$ cannot express the energy levels of the system directly since $E_C \ll E_J$ in the transmon and the PPQ. Therefore, the next step is to figure out the expression for the eigenenergy state $\ket{m}$ represented by the charge basis. We limit the range of summation to the subset of $\mathbb{Z}$, $\mathbb{S} = \{n:n\in\mathbb{Z},|n|\leq 50\}$, to express the bounded energy levels. Then, we obtain the $101\times 101$ idle Hamiltonian matrix $\hat{H}^{(\text{idle}, m)}$, represented by the eigenenergy basis, from the $101\times 101$ idle Hamiltonian matrix $\hat{H}^{(\text{idle}, n)}$ represented by the charge basis,

\begin{equation} 
  \hat{H}^{(\rm idle, m \it)} = \hat{V}^\dagger \hat{H}^{(\rm idle \it, n)} \hat{V},
\end{equation}
where $V$ is the change-of-basis matrix mapping from the charge basis to the eigenenergy basis. The other operators are also rewritten on the eigenenergy basis through the linear transformation $V$. Finally, we truncate the operators to $4 \times 4$ matrices such that the Hilbert space is now expressed as $\mathcal{H}$. All the operators of the transmon and PPQ follow this process and are represented by the eigenenergy basis.

\section{Two-level system Approximation}
\subsection{Transmon}
The transmon qubit is a superconducting circuit that behaves like an anharmonic oscillator. The Hamiltonian of a transmon can be expressed as,
\begin{equation}
    \hat{H}_T/\hbar = 4E_{C,T}\hat{n}_T^2 - E_{J,T} \cos \hat{\phi}_T.
\end{equation}
Note that we use $\hbar=1$. This can be approximated as a perturbated harmonic oscillator by a quartic potential well as follows:
\begin{equation}
    \hat{H}_T/\hbar \simeq 4E_{C,T}\hat{n}_T^2 - E_{J,T} \left(
        1 - \frac{\hat{\phi}_T^2}{2} + \frac{\hat{\phi}_T^4}{24}
    \right).
\end{equation}
Here, $\hat{H}_0/\hbar = 4E_{C,T}\hat{n}_T^2 - E_{J,T} \left(1 - {\hat{\phi}_T^2}/{2}\right)$ and $\hat{H}_1/\hbar = -(E_{J,T}/24)\hat{\phi}_T^4$ are the zeroth and first order Hamiltonians, respectively.
We define annihilation and creation operators for the transmon as,
\begin{align}
    \hat{b}_T &= A\hat{n}_T + iB\hat{\phi}_T,\\
    \hat{b}_T^\dagger &= A^*\hat{n}_T - iB^*\hat{\phi}_T.
\end{align}
Using the uncertainty principle of $[\hat{\phi}_T,\hat{n}_T]=i$, the number operator is expressed as,
\begin{align}
    \hat{b}_T^\dagger \hat{b}_T &= \left|A\right|^2\hat{n}_T^2 + \left|B\right|^2\hat{\phi}_T^2 + A^*B\\
    &= \frac{1}{\omega_0}\left(\hat{H}_0/\hbar + E_{J,T}\right) - \frac{1}{2}\\
    &= \frac{1}{\omega_0}\left(
        4E_{C,T}\hat{n}_T^2 + E_{J,T} \frac{\hat{\phi}_T^2}{2}
    \right) - \frac{1}{2},
\end{align}
where $\omega_0 = \sqrt{8E_{C,T}E_{J,T}}$ is the resonance frequency of the harmonic oscillator. Then, we can find the coefficients $A$ and $B$ as,
\pagebreak
\begin{equation}
    A = \sqrt{\frac{4E_{C,T}}{\omega_0}} = \frac{1}{\sqrt{2}}\left(\frac{8E_{C,T}}{E_{J,T}}\right)^{1/4},
\end{equation}
\begin{equation}
    B = \sqrt{\frac{E_{J,T}}{2\omega_0}} = \frac{1}{\sqrt{2}}\left(\frac{E_{J,T}}{8E_{C,T}}\right)^{1/4}.
\end{equation}
One can choose the degree of freedom of the phase $i^k$($k=0,1,2,3$) in the coefficients. Here, we choose $k=0$. Therefore, the second quantization of the transmon is given by,
\begin{equation}
    \hat{H}_T/\hbar = \sqrt{8E_{C,T}E_{J,T}}\left(\hat{b}_T^\dagger \hat{b}_T + \frac{1}{2}\right) - E_{J,T}- \frac{E_{C,T}}{12}\left(\hat{b}_T^\dagger - \hat{b}_T\right)^4,
\end{equation}
where,
\begin{align}
    \hat{n}_T &= \frac{1}{\sqrt{2}}\left(\frac{E_{J,T}}{8E_{C,T}}\right)^{1/4}\left(\hat{b}_T + \hat{b}_T^\dagger\right),\\
    \hat{\phi}_T &= \frac{-i}{\sqrt{2}}\left(\frac{8E_{C,T}}{E_{J,T}}\right)^{1/4}\left(\hat{b}_T - \hat{b}_T^\dagger\right).
\end{align}
The perturbated energy levels are given by,
\begin{equation}
    E_{m_T}^{(1)} = \bra{m_T}\hat{H}_1\ket{m_T} = -\frac{E_{C,T}}{12}\bra{m_T}(\hat{b}_T^\dagger - \hat{b}_T)^4\ket{m_T},
\end{equation}
We note the expansion of $(\hat{b}_T^\dagger - \hat{b}_T)^4$ below.
\begin{align*}
    (\hat{b}_T^\dagger - \hat{b}_T)^4 &= 
    \left(
        \hat{b}_T^2 + \hat{b}_T^{\dagger 2} - \hat{b}_T\hat{b}_T^\dagger - \hat{b}_T^\dagger\hat{b}_T
    \right)^2 \\
    &= 
    \left(
        \hat{b}_T^2 + \hat{b}_T^{\dagger 2} - 2\hat{b}_T^\dagger\hat{b}_T - 1
    \right)^2\\
    &=
    \left(
        \hat{b}_T^4 + \hat{b}_T^2\hat{b}_T^{\dagger 2} - 2\hat{b}_T^2\hat{b}_T^\dagger\hat{b}_T - \hat{b}_T^2
    \right.\\
    &\quad
    +\hat{b}_T^{\dagger 2}\hat{b}_T^2 + \hat{b}_T^{\dagger 4} - 2\hat{b}_T^{\dagger 3}\hat{b}_T - \hat{b}_T^{\dagger 2}\\
    &\quad
    -2 \hat{b}_T^\dagger\hat{b}_T\hat{b}_T^2 -2 \hat{b}_T^\dagger\hat{b}_T\hat{b}_T^{\dagger 2} +4 \hat{b}_T^\dagger\hat{b}_T\hat{b}_T^\dagger\hat{b}_T + 2 \hat{b}_T^\dagger\hat{b}_T\\
    &\quad
    \left.
            - \hat{b}_T^2 - \hat{b}_T^{\dagger 2} + 2\hat{b}_T^\dagger\hat{b}_T + 1
    \right)
\end{align*}
Then, the perturbated energy levels can be expressed as,
\begin{equation}
    E_{m_T}^{(1)} = -\frac{E_{C,T}}{12}\left(
        6m_T^2+6m_T+3
    \right).
\end{equation}
Thus, the Hamiltonian of the transmon approximated by a anharmonic oscillator is obtained as,
\begin{equation}
    \hat{H}_T/\hbar \simeq (\sqrt{8E_{C,T}E_{J,T}}-E_{C,T}) \hat{b}_T^\dagger \hat{b}_T - \frac{E_{C,T}}{2}\hat{b}_T^\dagger\hat{b}_T\left(\hat{b}_T^\dagger\hat{b}_T - 1\right).
\end{equation}
If we limit the basis states to the first two energy levels, the Hamiltonian can be expressed as,
\begin{equation}
    \hat{H}_T/\hbar \simeq \left(\sqrt{8E_{C,T}E_{J,T}}-E_{C,T}\right)\ketbra{m_T=1}{m_T=1}.
\end{equation}

\subsection{PPQ}
The Hamiltonian of a PPQ is given as,
\begin{equation}
    \hat{H}_P/\hbar = 4E_{C,P}\hat{n}_P^2 - E_{J,P} \cos 2\hat{\phi}_P.
\end{equation}
We focus on the dynamics of the PPQ in the odd parity. We can re-express the Hamiltonian for odd parity of Cooper pairs with $\hat{\mathcal{N}}_P$ and $\hat{\delta}_P$, the reduced charge and phase operators of a pair of Cooper-pairs, as follows:
\begin{equation}
    \hat{H}_P/\hbar = 4E_{C,P}\left(2\hat{\mathcal{N}}_P+1\right)^2 - E_{J,P} \cos \hat{\delta}_P.
\end{equation}
This Hamiltonian explains the dynamics of the PPQ in the odd parity. Here, we ignore the effect of the offset charge. It is reasonable because the dynamics of opposite-parity states are decoupled and each dynamics can be approximated by the same Hamiltonian\cite{Smith20}. Likewise the transmon, we can approximate the Hamiltonian of the PPQ as a perturbated harmonic oscillator by a quartic potential well as follows:
\begin{equation}
    \hat{H}_P/\hbar \simeq (\sqrt{32E_{C,P}E_{J,P}}-4E_{C,P}) \hat{b}_P^\dagger \hat{b}_P - 2E_{C,P}\hat{b}_P^\dagger\hat{b}_P\left(\hat{b}_P^\dagger\hat{b}_P - 1\right),
\end{equation}
where $\hat{b}_P$($\hat{b}_P^\dagger$) is the annihilation (creation) operators of the PPQ, respectively,
\begin{equation}
  \hat{b}_P = \frac{1}{\sqrt{2}}\left(\frac{32E_{C,P}}{E_{J,P}}\right)^{1/4}\hat{\mathcal{N}}_P
  + \frac{i}{\sqrt{2}}\left(\frac{E_{J,P}}{32E_{C,P}}\right)^{1/4}\hat{\delta}_P.
\end{equation}
Thus, the Hamiltonian of the PPQ reduced to TLS is given by,
\begin{equation}
    \hat{H}_P/\hbar \simeq \left(\sqrt{32E_{C,P}E_{J,P}}-4E_{C,P}\right)\ketbra{m_P=2}{m_P=2}.
\end{equation}

\section{Single-qubit gates}
\begin{figure*}[t]
  \centering
  \includegraphics*[width=1.0\linewidth]{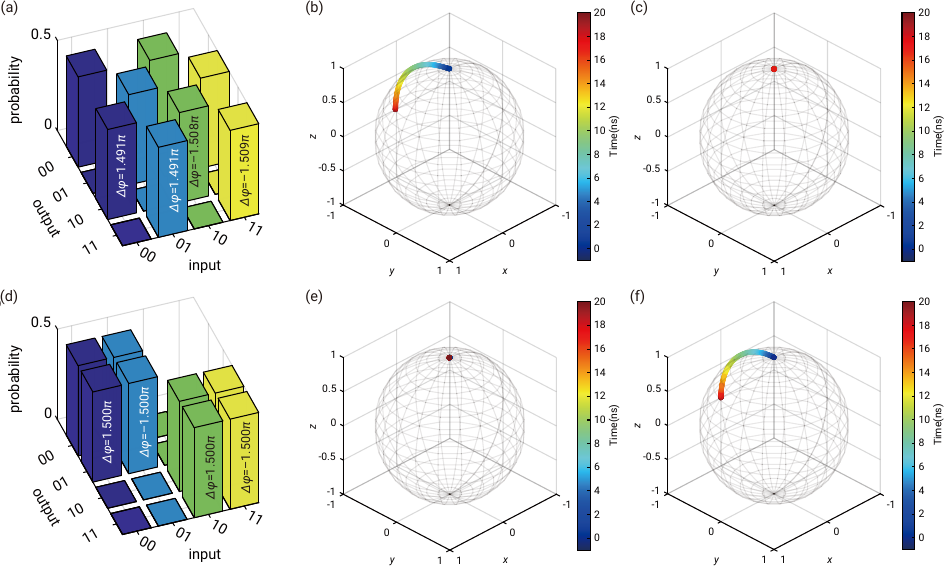}
  \caption{
    Numerical probabilities and the trajectory of the Bloch vectors for the basis states after applying the $R_X(\pi/2)$ gate. $\Delta\varphi$ denotes the relative phase for the state in the same column. The initial state for the simulation is $\ket{00}$.
    (a) State tomography simulated with the $R_{X,T}(\pi/2)$.
    The trajectory of the Bloch vectors for (b)the transmon and (c)the PPQ during the $R_{X,T}(\pi/2)$.
    (d) State tomography simulated with the $R_{X,P}(\pi/2)$.
    The trajectory of the Bloch vectors for (e)the transmon and (f)the PPQ during the $R_{X,P}(\pi/2)$.
  }
  \label{fig:1Q_Bloch}
\end{figure*}
\begin{table*}[t]
  \centering
  \caption{
      The pulse parameters of $R_X(\pi/2)$ gate and the average fidelity $F$ in the transmon-PPQ system. The pulse parameter vector $\mathbf{x}_\text{1Q}$ for the single-qubit gate is defined as $(f_q,T_G,\Omega_G,\gamma,\beta)$, where $f_q$, $T_G$, and $\Omega_G$ represent the frequency, time, and amplitude of the pulse, respectively. We set the initial phase of the pulse, $\gamma$, as zero for the $X$ rotation. $\beta$ is the DRAG coefficient.
  }
  \label{table:1Qgate_pulse_parameter}
  \begin{tabular*}{\textwidth}{@{\extracolsep{\fill}}ccccccc}
  \hline
  Gate             & $f_q$  & $T_G$ & $\Omega_G$ & $\gamma$ & $\beta$ & $F$    \\ \hline
  $R_{X,T}(\pi/2)$ & 2.8830 & 20    & -0.0154    & 0        & 0.3979  & 0.9996 \\ 
  $R_{X,P}(\pi/2)$ & 2.8470 & 20    & -0.0133    & 0        & 0       & 0.9995 \\ \hline
  \end{tabular*}
\end{table*}
We design single-qubit gates of the transmon and PPQ to construct a universal gate set. All single-qubit gates can be decomposed with a combination of $R_X(\pi/2)$ gates, modeled by using microwave voltage pulses, and $R_Z(\theta)$ gates, virtual Z(VZ) gates for frame rotations\cite{McKay17}. The voltage pulses denote the gate offset numbers $n_{g,T}(t)$ or $n_{g,P}(t)$.
The general form of the pulse is defined as a linear combination of sinusoidal functions with a time-varying envelope $\Omega_k(t)$:
\begin{equation}
  n_{g,i}(t) = \sum_{k}{\Omega_k(t)\cos(2\pi f_k t - \gamma_k)}.
\end{equation}
Here the index $i\in\{T,P\}$ denotes either the transmon or the PPQ in the transmon-PPQ system.
$f_k$ is the $k$-th pulse frequency in GHz, corresponding to the final frequency of the synthesized signal generated by a local oscillator and an arbitrary waveform generator (AWG). Since we utilize monochrome microwave signals, $f_k$ is fixed as a constant rather than a distribution. The pulse frequency determines the transition domain. For instance, if the pulse frequency is equal to the qubit frequency of the target qubit, the qubit states involved in the transition domain are $\ket{0}$ and $\ket{1}$. In this case, $f_k$ is the pulse frequency of the $R_X$ gate. If the pulse frequency is slightly detuned from this case, $Z$ rotation occurs owing to the off-resonant driving. The pulse amplitude $\Omega_k(t)$ is the signal generated by the AWG, stimulating the adiabatic process. It implies that designing the envelope which ensures proper adiabaticity is important for quantum gate design.

The pulse protocol for single-qubit gate $R_X(\pi/2)$ is defined as,
\begin{equation}
  n_{g,q}(t) = \Omega_G(t) \cos(2\pi f_q t - \gamma)
  + \Omega_{\text{DRAG},q}(t) \sin(2\pi f_q t -\gamma),
\end{equation}
where $q\in{T,P}$ indicates the transmon($T$) or the PPQ ($P$). We employ a Gaussian-shape envelope $\Omega_G(t)$ to design the single-qubit gates,
\begin{equation}
  \Omega_G (t) = \left\{
    \begin{array}{ll}
      \Omega_G \frac{\exp\left\{-\left(t-T_G/2\right)^2 / 2\sigma^2\right\}
      - \exp\left(-T_G^2/8\sigma^2\right)}
      {1 - \exp\left(-T_G^2/8\sigma^2\right)} & (0 \leq t < T_G)\\
      0 & \text{Otherwise}
    \end{array}
  \right.,
\end{equation}
where $f_q$ represents the qubit frequency of the target qubit. The qubit frequency of each qubit is given as $\omega_T = 2.883$ GHz, $\omega_P = 2.847$ GHz. $\sigma = T_G/4$ is the thickness of $\Omega_G(t)$.
This type of envelope can easily control a transmon-type superconducting qubit based on the Rabi oscillation.
We use Derivative Removal by Adiabatic Gate (DRAG)\cite{Chow10} to suppress the leakage error of the single-qubit gates,
\begin{equation}
  \Omega_{\text{DRAG},q}(t) = \beta \dot{\Omega}_G (t).
\end{equation}
The coefficient $\beta$ is proportional to the inverse of the anharmonicity of the qubit. We note that the DRAG pulse is advantageous only to the transmon in terms of gate fidelity. Thus, in this study, we do not apply the DRAG pulse for the single-qubit gates of the PPQ.

Table \ref{table:1Qgate_pulse_parameter} displays the pulse parameters and gate fidelities of the $R_{X,q}(\pi/2)$ gates. Our numerical results show that the transmon-PPQ system can perform single-qubit gates with average fidelities exceeding 0.999. This also implies that the transition between states having the same parity in the PPQ is numerically implemented by a high-fidelity gate operation whose gate time is similar to that of the single-qubit gate for the transmon. We visualize the simulation of these gates to provide a comprehensive understanding.
Figure \ref{fig:1Q_Bloch} illustrates the results of state tomography and time evolutions on a Bloch sphere for $R_{X,q}(\pi/2)$. In Figure \ref{fig:1Q_Bloch}a and \ref{fig:1Q_Bloch}d, $\Delta\varphi$ denotes the relative phase corresponding to the basis state. For instance, if the input state is $\ket{10}$, the output state is given as $(\ket{00}+e^{-1.508\pi i}\ket{10})/\sqrt{2}\approx (\ket{00}+i\ket{10})/\sqrt{2}$. Figure \ref{fig:1Q_Bloch}a and \ref{fig:1Q_Bloch}d show that the output states contain accurate phase information. We also investigate the time evolution of each qubit during the gate operations. Figure \ref{fig:1Q_Bloch}b(c) and \ref{fig:1Q_Bloch}e(f) show the trajectories of the transmon(PPQ) states. The trajectories of the Bloch vectors also confirm that the gates are properly designed.

\section{Gate Optimization}
We obtain the time evolution operator, the quantum gate, using the Hamiltonian matrix and the pulse parameters.
The time evolution operator $\mathcal{U}$ from time $t=0$ to $t=T$ is given by:
\begin{equation}
  \hat{\mathcal{U}}(T,0) = \mathcal{T} \exp{\left[
    -i \int_{0}^{T}{\hat{H}(t) dt}
  \right]}.
\end{equation}
Here, $\mathcal{T}$ is the time-ordering operator. We reproduce $\hat{\mathcal{U}}$ as the multiple time evolutions during the short time interval $\tau$. If $\tau$ is small enough, $\hat{H}(t)$ can be approximated to a constant matrix.
Thus, the time evolution operator $\mathcal{U}$ from the time $t=0$ to $t=T$ can be rewritten as the series of the time evolutions such that,
\begin{equation}
  \hat{\mathcal{U}}(T,0) \simeq \mathcal{T} \prod_{n=0}^{N}{\exp{\left[
    -i\tau \hat{H}(\{2n+1\}\tau / 2)
  \right]}}.
  \label{eq:time_evolution_numerical}
\end{equation}
We set $\tau$ as 1 ps.
However, these matrix exponential terms are expensive to compute using exact diagonalization method.
We calculate these terms efficiently using the second-order Suzuki-Trotter approximation\cite{Suzuki85}.
The total Hamiltonian $\hat{H}(t) = \hat{H}_0 + \hat{H}_1(t)$ is decomposed to the diagonal Hamiltonian $\hat{H}_0$ and the nondiagonal Hamiltonian $\hat{H}_1(t)$,
\begin{equation}
  \begin{array}{l}
    \hat{H}_0 = \hat{H}_T^\text{(idle)} + \hat{H}_P^\text{(idle)} + \hat{H}_R, \\
    \hat{H}_1(t) = \sum_{i\in\{T,P\}}\left(
      -8E_{C,i}n_{g,i}(t)\hat{n}_i
    \right) + \hat{H}_I.
  \end{array}
\end{equation}
We apply the approximation using the Hamiltonians decomposed by two parts,
\begin{equation}
  e^{
    -i\tau \hat{H}(\{2n+1\}\tau / 2)
  } \simeq
  e^{
    -i\tau \hat{H}_0/2
  }
  e^{
    -i\tau \hat{H}_1(\{2n+1\}\tau / 2)
  }
  e^{
    -i\tau \hat{H}_0/2
  }.
\end{equation}
Note that computing the matrix exponentials of $\hat{H}_1(t)$ is relatively cheap in terms of computation cost, due to the time-inedependency of all operators composing $\hat{H}_1(t)$.

\begin{figure}[t]
  \centering
  \includegraphics[width=1.0\linewidth]{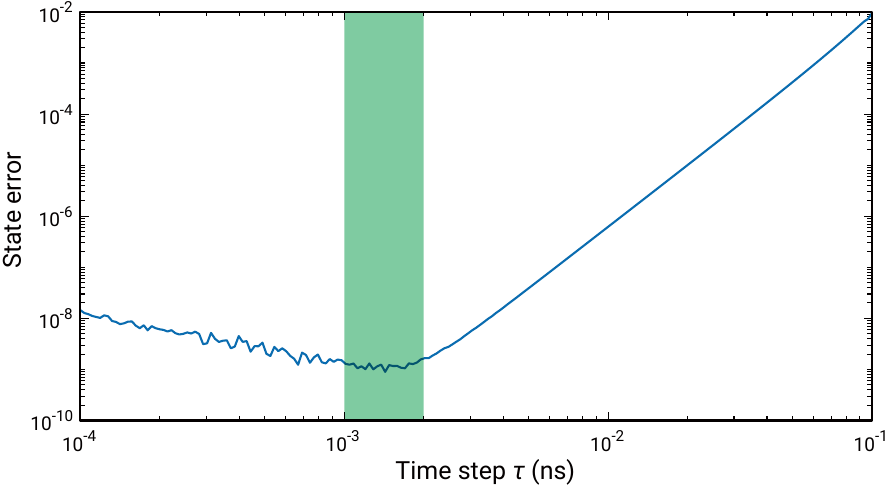}
  \caption{
    State error according to the time interval $\tau$. The difference between the exact diagonalization and the Suzuki-Trotter approximation is minimized in the green region.
  }
  \label{fig:time_error}
\end{figure}

Moreover, we can simulate almost the same dynamics computed by exact diagonalization using the second-order Suzuki-Trotter approximation. Figure \ref{fig:time_error} displays the infidelity between the evolved states calculated using exact diagonalization and the Suzuki-Trotter approximation. We simulate the free evolution over 100 ns, where the initial state is set by $\ket{++}$ in the transmon-PPQ system.
The time evolution operator $\hat{\mathcal{U}}_\text{ED(ST)}$ is calculated using exact diagonalization(Suzuki-Trotter approximation). We calculate the infidelity which we designate as `state error', $\mathcal{E}$, as follows:
\begin{equation}
  \mathcal{E} = 1 - \left|
    \bra{++}\hat{\mathcal{U}}_\text{ED}^\dagger
    \hat{\mathcal{U}}_\text{ST}\ket{++}
  \right|.
\end{equation}
We investigate the state error in the range of $\tau \in [10^{-4},10^{-1}]$ in Figure \ref{fig:time_error}. The state error decays as the time step becomes smaller and is minimized in the vicinity of $\tau = 10^{-3}$ ns. The error rather increases in $\tau < 10^{-3}$ ns. This may be due to the accumulated error during the calculation of the matrix exponentials. Thus, we show that the time interval $\tau$ is reasonably set for the simulation of the transmon-PPQ system.

We estimate the quantum gate $\hat{\mathcal{G}}$ using the fidelity $F$.
In this study, the gate fidelity for arbitrary pure states is approximated as the average fidelity of $\mathcal{N}=10^4$ arbitrary pure states as follows:
\begin{equation}
  F
  \simeq \sum_{n=1}^{\mathcal{N}}\left|\bra{\psi_n}
  \hat{\mathcal{G}}_{\text{ideal}}^\dagger
  \hat{\mathcal{G}}_{\text{pulse}}
  \ket{\psi_n}\right|,
\end{equation}
where $\hat{\mathcal{G}}_{\text{ideal}}$ denotes an ideal gate.
Here, we do not consider the time evolution involving leakage errors. That is, the average fidelity is estimated only by the computational basis.

We define a simple optimization problem to optimize the quantum gates.
\begin{equation}
  \min_{\mathbf{x}}\quad I = 1-F(\hat{\mathcal{U}}(\mathbf{x}))
\end{equation}
It is an optimization problem to minimize the gate infidelity $I$ subject to the pulse parameter vector $\mathbf{x}$ without any constraint. We employ the Nelder-Mead algorithm to solve this problem\cite{Nelder65}, because the gradient of the infidelity cannot be easily estimated in real quantum hardwares.

\end{document}